\documentclass[unnumsec,webpdf,contemporary,large]{oup-authoring-template}%

\graphicspath{{Fig/}}

\theoremstyle{thmstyleone}%
\theoremstyle{thmstyletwo}%
\theoremstyle{thmstylethree}%

\usepackage{supertabular}
\usepackage{microtype}
\usepackage{multirow}
\usepackage{graphicx}
\usepackage{booktabs} 
\usepackage[figuresright]{rotating} 
\usepackage{makecell} 
\usepackage{hyperref}
\usepackage{natbib}
\setcitestyle{numbers,square,sort,comma}
\usepackage{xcolor}
\usepackage{amssymb}
\usepackage{amsfonts}
\usepackage{longtable}
\usepackage{mathrsfs}
\usepackage{marvosym}
\usepackage{threeparttablex}
\usepackage{appendix}

\usepackage{caption}
\DeclareCaptionLabelFormat{mylabel}{\textbf{#1 #2.}\hspace{0.8ex}}
\begin{document}

\journaltitle{Journal Title Here}
\DOI{DOI HERE}
\copyrightyear{2021}
\pubyear{2021}
\access{Advance Access Publication Date: Day Month Year}
\appnotes{Paper}

\firstpage{1}


\title[Protein-RNA interaction prediction with deep learning: Structure matters]{Protein-RNA interaction prediction with deep learning: Structure matters}

\author[1,$\dagger$]{Junkang~Wei}
\author[2,$\dagger$]{Siyuan~Chen}
\author[1,$\dagger$]{Licheng~Zong}

\author[2,$\ddagger$]{Xin~Gao}
\author[1,3,$\ddagger$]{Yu~Li}

\authormark{Junkang~Wei et al.}

\address[1]{\orgdiv{Department of Computer Science and Engineering (CSE)}, \orgname{The Chinese University of Hong Kong (CUHK)}, \orgaddress{\postcode{999077}, \state{Hong Kong SAR}, \country{China}}}

\address[2]{\orgdiv{Computational Bioscience Research Center (CBRC)}, \orgname{King Abdullah University of Science and Technology (KAUST)}, \orgaddress{\postcode{23955-6900}, \state{Thuwal}, \country{Saudi Arabia}}}

\address[3]{\orgname{The CUHK Shenzhen Research Institute}, \orgaddress{\street{Hi-Tech Park}, \postcode{518057}, \state{Shenzhen}, \country{China}}}

\corresp[$\dagger$]{These authors contributed equally. \\}

\corresp[$\ddagger$]{Corresponding author. \href{email:xin.gao@kaust.edu.sa}{xin.gao@kaust.edu.sa  } \href{email:liyu@cse.cuhk.edu.hk}{liyu@cse.cuhk.edu.hk}}

\received{Date}{0}{Year}
\revised{Date}{0}{Year}
\accepted{Date}{0}{Year}



\abstract{Protein-RNA interactions are of vital importance to a variety of cellular activities. Both experimental and computational techniques have been developed to study the interactions. Due to the limitation of the previous database, especially the lack of protein structure data, most of the existing computational methods rely heavily on the sequence data, with only a small portion of the methods utilizing the structural information. Recently, AlphaFold has revolutionized the entire protein and biology field. Foreseeably, the protein-RNA interaction prediction will also be promoted significantly in the upcoming years. In this work, we give a thorough review of this field, surveying both the binding site and binding preference prediction problems and covering the commonly used datasets, features, and models.  We also point out the potential challenges and opportunities in this field. This survey summarizes the development of the RBP-RNA interaction field in the past and foresees its future development in the post-AlphaFold era.}


\maketitle
\section{Introduction}\label{sec:introduction}

Protein-RNA interactions are involved in a variety of cellular activities, such as gene expression regulations \citep{RN1444}, post-transcriptional regulations \citep{RN1427}, and protein synthesis \citep{RN1432}. Perturbation of such interactions can lead to fatal cellular dysfunction and diseases \citep{RN1391}. Owing to their importance, researchers have made significant efforts to understand the interactions \citep{RN1374} and the related molecular mechanism behind the processes \citep{RN1398,RN1446}. Due to the difficulty to perform high-throughput structural biological experiments in the last century, the progress of this field was slow \citep{RN1437}. However, with the development and advancement of high-throughput assays, such as the \textit{in vivo} RIP-seq \citep{keene2006rip} and CLIP-seq \citep{ule2005clip}, and the \textit{in vitro} RNACompete \citep{RN820} and HT-SELEX \citep{RN827}, we have witnessed the significant progress of this field as well as the large amount of accumulated data \citep{RN1427}. Computational methods emerge to analyze the data and accelerate the discovery \citep{RN1371,RN1369,RN780,RN1383,RN1400,RN1410}. 

Similar to the experimental techniques, which can be divided into the structure-based methods and the assay-based methods, the computational methods can also be classified into two categories, either predicting the RNA binding sites on the protein surface \citep{RN1431,RN1432} or modeling the preferred RNA sequences of an \textcolor{black}{RNA-Binding Protein} (RBP) \citep{RN1396}. In the first category, people essentially resolve a binary classification problem. Given the protein, researchers want to predict whether it is an RBP, and if it is an RBP, at which amino acids it can interact with an RNA. In the latter one, given a protein with the high-throughput assay experimental data, people extract the frequency of each nucleotide at each position on the preferred RNA sequences, using \textit{k}-mer models \citep{lee2015method}, \textcolor{black}{Position Weight Matrix} (PWM) models \citep{RN1444}, or deep learning models \citep{RN1427}. If the computational method targets on genome-wide prediction, sometimes, it is also referred as the binding sites prediction on RNAs \citep{pan2018predicting,li2017deep}, which may cause confusion to the readers. In the rest of the paper, binding sites prediction refers to predicting the RNA binding sites on the protein surface, while the binding preference prediction refers to predicting the protein binding preference against RNA sequences. On the other hand, as both of the two main research directions are protein-centric \citep{RN1391}, which means that there is intrinsic relation between the two research topics, researchers are also trying to predict both information simultaneously with a unified deep learning method \citep{RN1400}. 
\begin{figure*}[htp]
  \centering
  \includegraphics[width=\textwidth]{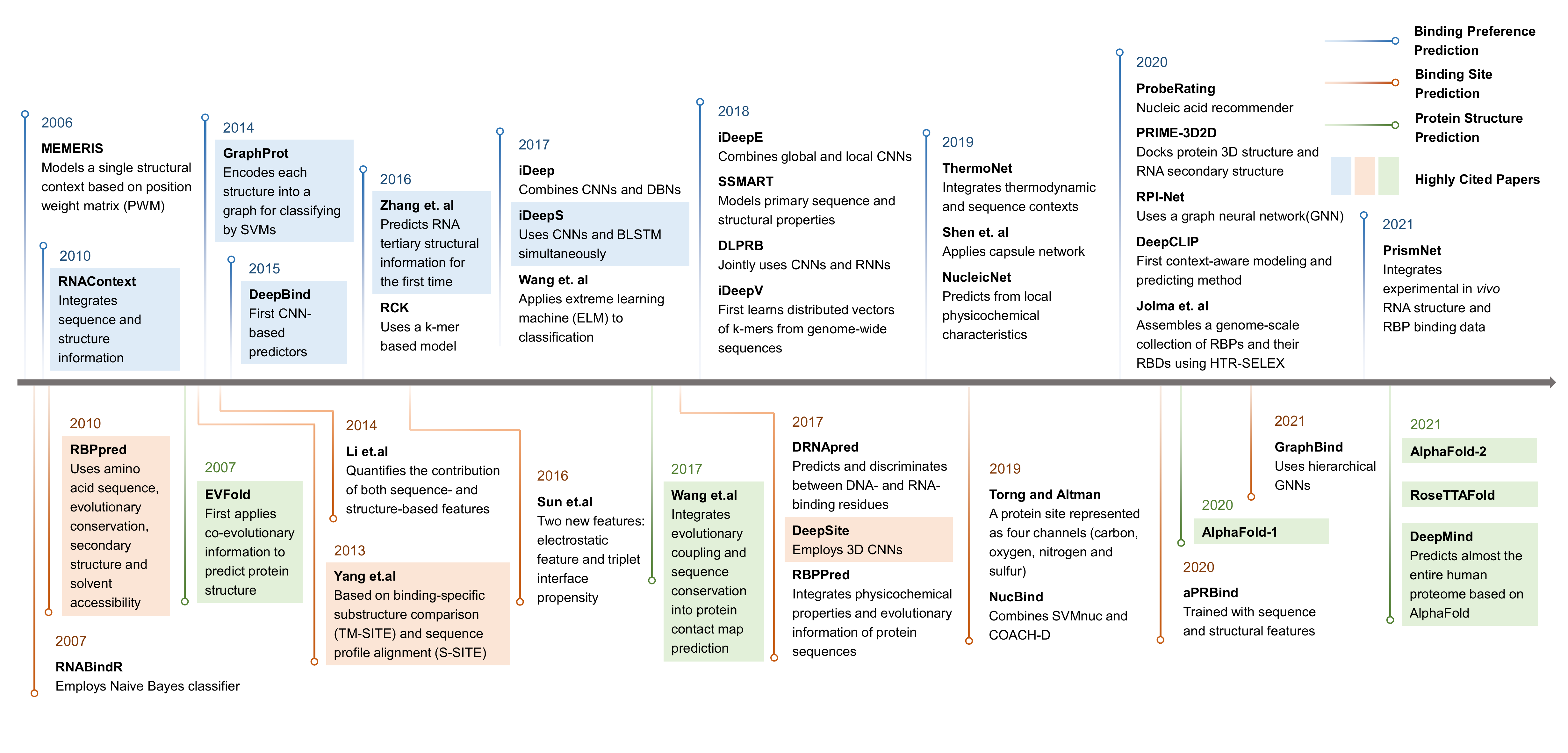}
  \caption{An overview of important works related to binding site and binding preference prediction. \textcolor{black}{Protein structure prediction methods are also included because of their rising importance in interaction prediction. The three categories are represented by different colored lines in chronological order. Highly cited papers are highlighted by corresponding colored boxes. Following the significant progress in the past years, this field will embrace great advancement in the upcoming years.}}
    \label{reviews}
\end{figure*}

Since the first computational method was proposed to tackle the interaction between RNA and protein specifically \citep{RN1439}, a number of algorithms have been developed to handle the problems \citep{RN1431,RN1432,RN1382,RN1388,RN1387}. They can be divided into the following categories. Firstly, based on the assumption that similar structures may have similar function, people have used the template-based method to predict the binding sites \citep{yang2013protein,chen2014identifying,wu2018coach,xie2020prime} and the binding preference \citep{zheng2016template}. Although such methods can perform well on queries with homologs, they have difficulty in handling new sequences without homologs \citep{senior2020improved}. Secondly, people combine hand-crafted features, which will be discussed in the next paragraph, with shallow-learning methods, such as \textcolor{black}{Support Vector Machine} (SVM) \citep{RN1379,RN1433,zhang2017rbppred,su2019improving}, logistic regression \citep{kazan2010rnacontext,RN1421,hiller2006using,RN789}, and random forest \citep{RN1418,RN1425}, to investigate the topic. The commonly used \textit{k}-mer models \citep{RN1421} and PWM models \citep{kazan2010rnacontext} are classified into this category, because they are usually combined with logistic regression. Notice that this category of methods is still under active development \citep{zhang2017rbppred,su2019improving}, even after the surge of deep learning, because it is difficult to represent and encode the raw structural information, which will be discussed in detail in this paper. 
The last category is the deep learning-based methods \citep{RN1427,RN1400,RN1369}, which have been very popular in recent years. With such models, people only need to input the raw representation of the proteins or RNAs, and let the models learn and extract useful information by themselves. However, the transparency and interpretability of the models are usually questioned \citep{li2019deep}.

Within the above algorithms, people have been using various features, including the ones from both proteins and RNAs. Regarding the protein features, researchers have developed representations from sequences, such as sequence one-hot encodings \citep{RN1432}, \textcolor{black}{Position-Specific Scoring Matrix} (PSSM) \citep{su2019improving,RN1440}, and conservation entropy derived from PSSM. The physicochemical properties \citep{chen2008predicting}, including hydrophobicity, electrostatics, and atom types, are also helpful. Although the individual local protein structural information, such as residue propensity and solvent accessibility, has been adopted for a while \citep{RN1431}, recently, researchers have shown that directly using the comprehensive local structural encoding can significantly improve the model's performance  \textcolor{black}{\citep{RN1400,torng2019high}}. For example, people have used voxels \citep{torng2019high} and graphs \citep{RN1370} to encode the protein 3D structures. In terms of the RNA features, the logic is similar to the protein ones. Regarding the sequence features, people have been using the sequence one-hot encodings \citep{gronning2020deepclip,RN1369,RN1427}, \textit{k}-mer models \citep{RN1421}, and PWM \citep{kazan2010rnacontext,RN1421}. However, unlike the protein secondary structures, RNA secondary structural information has been significantly emphasized, including both the predicted RNA secondary structures and the \textit{in vivo} structure profiles \citep{hiller2006using,RN1433,kazan2010rnacontext,RN1369}. Meanwhile, the tertiary structures are also shown to be very important \citep{RN1426}.
Despite the large variety of existing features, unfortunately, people have not taken full advantage of them for the following two reasons. Firstly, in the binding site prediction, people usually only consider the protein information, while in the binding preference prediction, people usually only consider the RNA information. \textcolor{black}{Similar to Drug-Target Interaction (DTI), the interactions between RNAs and proteins include at least two molecules, and using information from only one side could lead to inferior performance \citep{li2020monn, luo2017network,zheng2020predicting}.} Secondly, the RNA and protein structural information has not been fully utilized as well, mainly due to the limitation of previous structure prediction methods and the unsatisfactory structure encoding methods.

\begin{figure*}[htp]
  \centering
  \includegraphics[width=\textwidth]{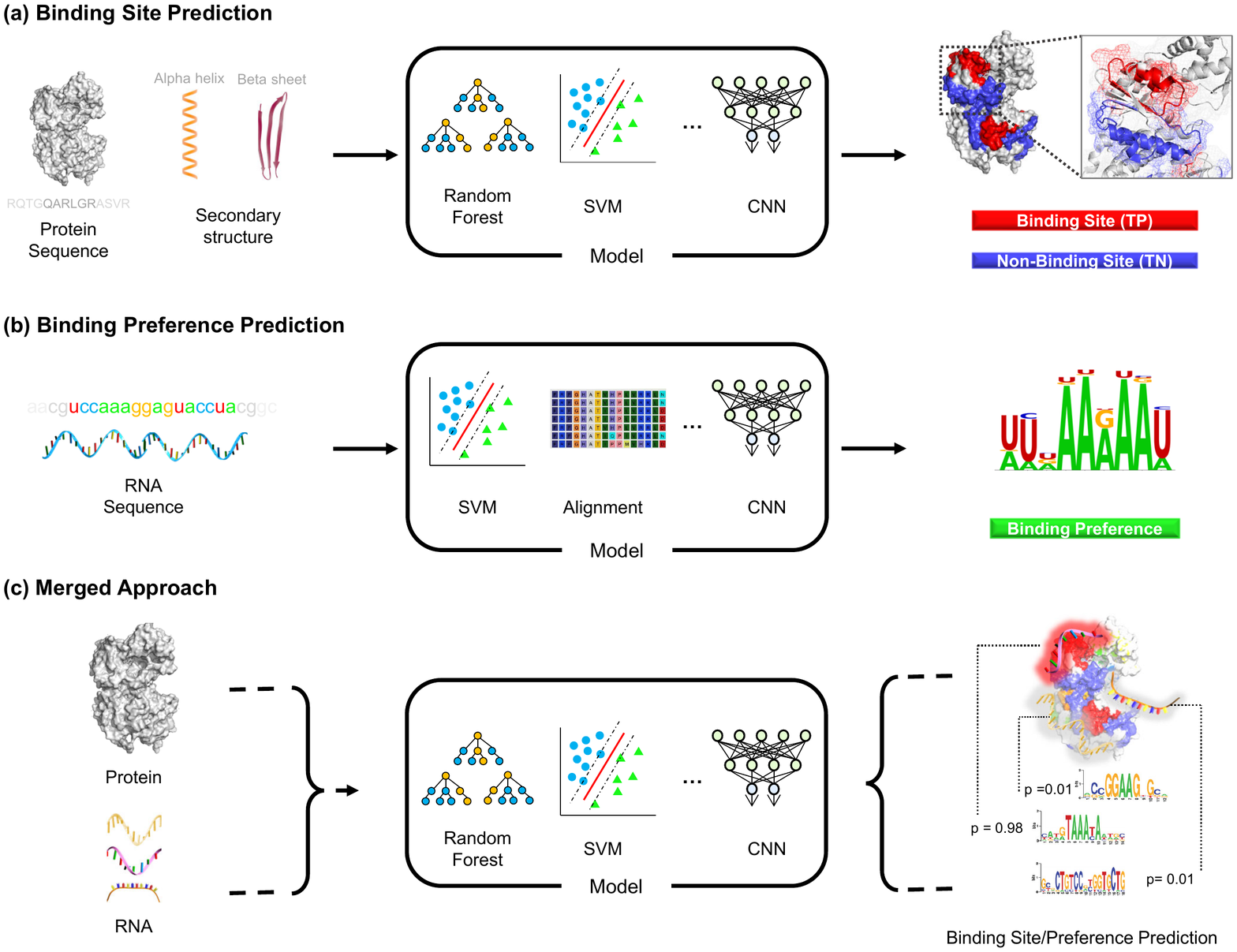}
  \caption{The different paradigms of studying the interactions between proteins and RNAs. \textbf{a}. Binding site prediction. Given the protein information, people predict which locations on the protein surface are the binding sites for RNAs. \textbf{b}. Binding preference prediction. For a given protein, the researchers have already determined the RNA sequences that can bind to the protein by experiments. Here, the models learn the statistical information from the input RNA sequences as the binding preference of that specific protein against RNAs. 
  \textbf{c}. For studying the interaction more comprehensively, it is more desirable to consider the protein and RNA information, including both the sequence and structural information, simultaneously and predict both binding sites and binding preference.
  }
    \label{2_problem}
\end{figure*}

In recent years, we have witnessed the significant improvement of both the structure determination methods \citep{yip2020atomic} and prediction methods \citep{marks2011protein,wang2017accurate,wang2019predmp,senior2020improved,jumper2021highly,Baekeabj8754,tunyasuvunakool_highly_2021}. Considering the success of the previous computational methods targeting protein-RNA interaction prediction based on structural information, it is foreseeable that researchers will make significant progress in this field (Figure \ref{reviews}). Given that, we review this field thoroughly in this paper, emphasizing the structural information. In this work, we also consider the protein-RNA interaction binding site and binding preference prediction simultaneously for the first time, considering their intrinsic relationship. 
We notice that there are some existing related reviews focusing on different aspects of this problem. More specifically, Pan et al. \citep{pan2019recent}; Yan and Zhu et al. \citep{RN1382}; Sagar and Xue et al. \citep{RN1388} list the recently developed deep learning tools for predicting binding preference. Trabelsi et al. \citep{RN1396} evaluates the performance of different deep learning models on predicting the binding preference. Yan et al. \citep{RN1432}; Si et al. \citep{RN1430}; Miao and Westhof \citep{RN1431} list and evaluate the tools for predicting binding sites on protein, while all the involved methods were developed before 2014, which means that the deep learning methods are not included. Hafner et al. \citep{RN1371}; Ramanathan et al. \citep{RN1391}, Licatalosi et al. \citep{RN1380}; Corley et al. \citep{RN1374} summarize the related biological experimental techniques to study the interactions as well as the biological insights and mechanism behind the interactions. \textcolor{black}{More recently, Jamasb et al. \citep{jamasb2021deep} concluded the computational methods for protein-protein interaction site prediction with deep learning approaches. Also, the work of Day et al. \citep{day2020message}, namely Message Passing Neural Processes (MPNPs), successfully improved the performance of the node classification task in the protein-protein interaction site prediction problem. It uses the protein structural data as the interacting residue graph, which thrives at lower sampling rates.} Our work, which unifies two intrinsically related computational problems and highlights the importance of structural information, can provide new insights into the topic. 
Table \ref{review_paper_table} summarizes the main focuses of different review papers.

This paper is organized as follows. In Section \ref{sec:problems}, we give a clear description of the computational problems related to the interaction between proteins and RNAs. From Section \ref{sec:datasets} to Section \ref{sec:evaluation}, we review each component of the computational methods targeting against the above problems, including datasets (Section \ref{sec:datasets}), features (Section \ref{sec:features}), models (Section \ref{sec:models}), and model evaluation (Section \ref{sec:evaluation}). In Section \ref{sec:challenges}, we provide a thorough review on the challenges and opportunities in this field. Although we emphasize on the importance of structural information to the interaction, for the completeness of this review, we also mention the methods only utilizing sequence encoding.




\begin{table*}[thp]

    \caption{Summary and comparison of the existing reviews on the studies of protein-RNA interaction. \textcolor{black}{Sorted by the published year, the reviews are divided into different categories based on their main focuses: CLIP, RNA-binding sites, 3D structural information, DNA-binding specificity, RNA-protein interaction data, and RNA-binding preferences.}
    }
    
    \setlength{\tabcolsep}{2mm}
    
    \begin{tabular}	{p{2cm}<{\centering} || p{2cm}<{\centering}  p{4cm}<{\centering}p{7.4cm}<{\centering}}
    \hline
    Paper & Year& Journal & Main Focus \\
    \midrule
    \citep{RN1434} & 2012 & Nature Review Genetics & State-of-the-art Ultraviolet \textcolor{orange}{Crosslinking and Immunoprecipitation(CLIP)} \\[18pt]
    
     \citep{RN1408} & 2018 & Molecular Cell & Rationale for each step in \textcolor{orange}{CLIP protocol and discuss the impact of variations technologies}\\[18pt]
    
    \citep{RN1398} & 2019 & Nucleic Acids Research & Assessment of RNA SS and \textcolor{orange}{CLIP} in detail\\[18pt]
    
    \citep{RN1371} & 2021 & Nature Reviews Methods Primers & Prospect of integrating data obtained by  \textcolor{orange}{CLIP}\\[18pt]
    \cline{2-4}


    \citep{RN1431} & 2015 &PLOS Computational Biology & Comprehensive assessment  on \textcolor{blue}{RNA-binding sites prediction} from multiple web servers, \textcolor{blue}{datasets}, and protein-nucleic acid complexes \\[18pt]
    
    \citep{RN1430} & 2015 & International Journal of Molecular Sciences & Computational approaches for RBPs and \textcolor{blue}{RNA-binding sites} prediction \\[18pt]

    \cline{2-4}
    
    \citep{RN1422} & 2016 & Biophysical Reviews &  \textcolor{magenta}{3D structure}  of protein–RNA complexes at the atomic resolution \\[18pt]

     \citep{RN1413} & 2017 & Nature Reviews Molecular Biology & The coupling of RNA modifications and \textcolor{magenta}{structures describe RNA–protein interactions} at different steps of the gene expression process \\[18pt]

    \citep{RN1407} & 2018 & Genes & Computational methods for macromolecular docking and for scoring \textcolor{magenta}{3D structural} models  of RNP complexes\\[18pt]

    \cline{2-4}

    \citep{RN1444} & 2013 & Nature Biotechnology & Systematical comparison of protein's \textcolor{olive}{DNA-binding specificity}\\[18pt]
    
     \citep{RN1432} & 2016 & Briefing in Bioinformatics & \textcolor{olive}{RNA-} or \textcolor{olive}{DNA-binding residues} from protein sequences\\[18pt]

    \citep{RN1396} & 2019 & Bioinformatics & Deep learning architectures for predicting \textcolor{olive}{DNA- and RNA-binding specificity} \\[18pt]
    
     \cline{2-4}

    \citep{RN1418} & 2015 & Briefings in Functional Genomics & Integrating \textcolor{violet}{RNA–protein interaction data} with observations of post-transcriptional regulation \\[18pt]

    \citep{RN1394} & 2019 & Journal of Biological Chemistry & Statistical inference and machine-learning approaches for RBPs prediction,  analysis of large-scale \textcolor{violet}{RNA–protein interaction datasets} \\[18pt]
    
    \citep{RN1388} & 2019 & Protein and Peptide Letters & Computational predictors for \textcolor{violet}{RNA-protein interaction} in the aspects of data, prediction, and input features \\[18pt]

    \citep{RN1380} & 2020 & Wiley Interdisciplinary Reviews:RNA &  \textcolor{violet}{RNA interactions with proteins} and techniques measuring the kinetic dynamics of RNA–protein interactions \textit{in vitro}\\[18pt]
    
    \citep{RN1391} & 2019 & Nature Methods & Comparison between \textcolor{violet}{RNA-centric and protein-centric} experimental methods\\[18pt]
    
     \citep{RN1374} & 2020 & Molecular Cell & \textcolor{violet}{Protein-RNA molecular interactions} \& Software availability\\[18pt]

     \cline{2-4}

    \citep{RN1382} & 2020 &IEEE Access & Machine learning and deep learning approaches focusing on \textcolor{teal}{RNA binding preference}\\[18pt]

    \citep{pan2019recent} & 2020 & Wiley Interdisciplinary Reviews: RNA & Prediction of RNA–protein interaction pairs and  \textcolor{teal}{RBP binding preference}\\[18pt]
    

  \bottomrule
  \end{tabular}
  \label{review_paper_table}\\
\end{table*}

\section{Computational problems for protein-RNA interaction}
\label{sec:problems}
In this section, we are going to introduce the two kinds of computational problems related to the interaction between proteins and RNAs in detail. As discussed in the Introduction, we refer to the first one as the binding sites prediction and the second one as the binding preference prediction. We summarize the paradigms in Figure \ref{2_problem}.

\subsection{\textbf{Binding sites prediction}}
\label{sub:sites}
This problem is related to the first problem that people want to know when investigating the protein and RNA interaction. Given a protein, we first want to know whether this protein is an RBP or not. If it is not an RBP, we could stop here and save the computational resources for other proteins. If the protein is an RBP, people further want to know which amino acids on the protein sequence can potentially interact with RNAs, which is related to the function of the protein. In other words, researchers want to predict the binding sites and binding positions on the protein surface for RNAs. 

Usually, for this problem, people only consider the information from the protein side. The  input is a protein, with either the sequence information or the structure information, or both. Then, researchers extract some features or define certain scoring functions with the above information. A machine learning model or an alignment-based method will thus be developed accordingly with an annotated database. The outputs are binary predictions, either at the protein level or the amino acid level. Usually, the methods based on structure have better performance on this problem than the sequence-based methods \citep{RN1431}, as the local structure can determine whether the protein is accessible for interaction with other molecules.

\subsection{\textbf{Binding preference prediction}}
In this computational problem, we want to know more information about the interaction from the RNA side. The interaction involves two molecules, a protein, and an RNA. In Section \ref{sub:sites}, we have investigated it from the protein side, determining which amino acids can potentially interact with RNAs. In this problem, we study which RNAs can interact with a certain protein. If we describe the problem from the protein aspect, we want to know the binding preference of the protein against RNAs.

Although we want to predict the binding preference of an RNA-binding protein, seldom would researchers include the protein information in the prediction model. Usually, the training data are a set of RNA sequences or RNA secondary structures, which are proved to interact with a protein. Then, a machine learning model or a statistical motif model will be constructed based on the data. The inputs of these models are RNA features, and the models will predict whether they can interact with the protein. Notice that, in these models, people do not use the protein information explicitly. Instead, people believe a large amount of training RNA sequences can describe the target protein implicitly. However, recent studies \citep{RN1400,xie2020prime} show that the protein information can be used directly to predict the interaction preference, even without the high-throughput assay data.


\section{Datasets for building the models}
\label{sec:datasets}
After defining the computational problems, we need to prepare the related data, which are the foundation for building computational models to resolve the above problems. The data can be divided into two categories, either the protein/RNA sequence data or the structure data. In this section, we give an overview of the data and the related databases. We also summarize the datasets in Table \ref{dataset_table}.

\subsection{\textbf{Sequence datasets}}
The protein sequences are usually used for predicting the binding sites, while the RNA sequences are used for predicting the binding preference. The techniques to sequence proteins are very mature, and the resulted data are stored in UniProt\footnote{https://www.uniprot.org}, which is one of the most famous databases in bioinformatics. 

The techniques to investigate the proteins' binding preference against RNAs include the \textit{in vivo} RIP-seq \citep{keene2006rip} and CLIP-seq \citep{ule2005clip}, and the \textit{in vitro} RNACompete \citep{RN820} and HT-SELEX \citep{RN827}. Although their experimental techniques and protocols are very different, the basic principles are the same, that is, to identify and isolate RNAs that a protein can interact with and then sequence those RNAs. Consequently, the outputs and the data from those experiments are RNA sequences. As this review does not focus on the experimental techniques, we refer the readers to the related reviews in case the readers are interested in them \citep{RN1398}. 

In Table \ref{dataset_table}, we list the related datasets. The doRiNA \citep{anders2012dorina} contains 24 experiments of 21 RBPs, which are determined by experimental protocols including PAR-CLIP (Ago/EIF2C1-4, IGF2BP1-3, PUM2, Ago2-MNase, ELAVL1, ELAVL1-MNase, ELAVL1A, ESWR1, FUS, TAF15, MOV10) and CLIP-seq (TIAL1, Ago2, ELAVL1, eIF4AIII, SRSF1). \textcolor{black}{On the other hand, iCount created the iCLIP dataset for 17 RBPs with 5996 binding sites. iONMF \citep{stravzar2016orthogonal} analyzed the data from iCount and doRiNA, building a unified dataset, which has been widely used in different models including iDeepE \citep{pan2018predicting}, deepnet-rbp \citep{pan2018learning}, deepRAM \citep{RN1396}.}

AURA 2 \citep{dassi2014aura} collects the \textcolor{black}{Untranslated Regions (UTRs) in mRNA sequences } of 67 RBPs with 502,178 binding sites. Within the dataset,  the number of binding sites for different RBPs is variant. \textcolor{black}{To eliminate bias from imbalance positive sample distribution, iDeepE constructed RBP-47, removing 20 RBPs with less than 2000 positive sequences.} However, the RBP-47 only provides the positive UTRs sequence. \textcolor{black}{For generating the negative sample, the RBP-47 selects the UTRs from other RBPs, excluding the binding sites in the target RBPs. It is different from the strategy of doRiNA, which generates the negative samples by selecting random sites excluding positive binding sites in the same gene. Intuitively, the doRiNA's tactic would be more rational and have a lower possibility of including false-negative samples. Theoretically, the CLIP-seq experiments detect regions as the binding sites of a gene and the other regions as unbinding sites, which means that experiments have verified the negative samples.} 

CLIPdb \citep{yang2015clipdb} is a database of various high-resolution binding sites for RBPs, collecting from published CLIP-seq data. It contains manually curated annotations from CLIP-seq studies across different organisms with 395 CLIP-seq samples for 111 RBPs. \textcolor{black}{In addition, CLIPdb also provides genome-wide binding sites for each dataset by a unified analysis. The resulted high-resolution binding site data from a large number of RBPs will benefit investigations on the coordination and competition of RBP binding mechanism. Because the binding sites of RBPs are identified by CLIP-seq and well-annotated in CLIPdb, its \textcolor{black}{negative sampling} setting is similar to that of \textcolor{black}{doRiNA}.}

\subsection{\textbf{Structure datasets}}

\textbf{Protein structure}: For the protein structure, the most comprehensive database is Protein Data Bank (PDB)\footnote{https://www.rcsb.org}. Although the database does not contain the structure of all the RNA-binding proteins and some parts of the RNAs may not be very clear, most of the existing structure datasets are extracted from structures of protein-RNA complexes from PDB \citep{sussman1998protein}. Generally, the criterion of the amino acid in the protein being considered as RNA-binding in a co-crystal complex, is that at least one of its backbone atoms or side chains are within a certain distance from atoms of the RNA. Specifically, both 3.5Å and 5.0Å are the usual threshold \citep{RN1432}. 

\textcolor{black}{\textcolor{black}{Nucleic Acid–Protein Interaction Database} (NPIDB) \citep{RN813} collects structural information of all the DNA-protein and RNA-protein complexes available from PDB. The dataset followed the classification by the binding nucleic acids such as RNA (668), DNA (1671), RNA \& DNA (504).} \textcolor{black}{On the other hand, ccPDB \citep{agrawal2019ccpdb} provides a dataset of DNA/RNA-interacting proteins, including 417 DNA binding proteins and 282 RNA binding proteins, and identifies their DNA/RNA-interacting residues. In addition, ccPDB collects nucleotide-protein interactions such as ATP-, GTP-, NAD-, FAD-protein interactions, which may have the parallel physicochemical mechanism with RNA-protein interaction.} RNA\_T dataset \citep{RN1432} is also a benchmark dataset collected from PDB, which consists of 981 RNA-binding protein chains with the distance cutoff of 3.5Å(985 for 5Å). \textcolor{black}{To alleviate the effect of chain replicates induced by strand truncation, the authors establish a dataset by removing chains with high sequence and structural similarities. The resulted dataset contains 175 representative and non-redundant RNA-binding protein chains.}

\textcolor{black}{Meanwhile, homologous protein structures may cause bias in modeling.} NucleicNet \citep{RN1400} has defined two homologous redundancy, internal redundancy and external redundancy. The internal redundancy is that multiple copies of the same RNA-binding protein chain can exist within the same PDB entry due to the formation of homo- or hetero-multimeric complexes. The external redundancy is that homologous chains are shared across different PDB entries and dedicated to different binding RNA sequences. These redundant RNA-binding samples, sharing the homologous chains common in RNA-binding configurations and physicochemical environments, would introduce bias to the evaluation and cause the overstated generalizability power of the model. To remove the internal redundancy, the authors retain the best locally resolved component and discard the other homologous protein and RNA. For the external redundancy, PDB entries are clustered into groups where each entry is linked with others that share at least one RNA-binding chain with cutoff=90\% BLASTClust sequence homology \citep{earl2007compare}. For each cluster, the PDB entry with the best resolution is selected, turning the 483 valid PDB entries into 158 clusters. The authors select one representative entry for each cluster.

\textcolor{black}{With the appearance of AlphaFold, Jumper et al.} \citep{jumper2021highly} provides AlphaFold Protein Structure Database, which contains 23,391 protein structures (\textit{Homo sapiens}) and covers 98.5\% of human proteome. Although it is a method of \textit{ab initio} protein structure prediction, AlphaFold can already achieve a similar prediction accuracy and resolution as Cryo-EM on some proteins. This means structures of RBPs that have not been successfully resolved by experimental approaches may have already been predicted accurately by AlphaFold.

\textbf{RNA secondary structure}: \textcolor{black}{Although most of the developed binding preference prediction methods only utilize the predicted secondary structure, such as RNAstructure \citep{reuter2010rnastructure} or SPOT-RNA \citep{singh2019rna}, to improve the prediction performance, there are datasets containing the experimentally determined RNA structures and \textit{in vivo} profiles.} Sun et al. \citep{RN1369} introduce icSHAPE \citep{flynn2016transcriptome} to characterize the single- and double-stranded regions of RNAs, which is crucial information to protein-RNA interaction. \textcolor{black}{Recently, RNA Atlas of Structure Probing (RASP) \citep{li2021rasp} collects transcriptome-wide RNA secondary structure probing data through 18 experimental methods such as DMS-seq, SHAPE-Seq, SHAPE-MaP, and icSHAPE, \textit{etc}.}

Intuitively, the experimental and well-annotated RNA secondary structure provide precise and informative input to modeling. For instance, bpRNA \citep{danaee2018bprna} collects 102,318 known secondary structures from 7 different databases, including Comparative RNA Web Site \citep{cannone2002comparative}, tmRNA Database \citep{zwieb2003tmrdb}, Signal Recognition Particle Database \citep{rosenblad2003srpdb}, Sprinzl tRNA Database, RNase P Database \citep{brown1998ribonuclease}, RNA Family Database \citep{griffiths2003rfam} and PDB. Besides, bpRNA introduces a novel annotation tool to parse complex pseudoknot-containing RNAs with 7 annotations, such as stems, internal loops, bulges, multi-branched loops, external loops, hairpin loops, and pseudoknots. \textcolor{black}{Furthermore, bpRNA offers a high-quality subset of the database with sequence similarity lower than 90\% identity, which helps the model solve the issue of training data replicates.}

\begin{table*}[thp]
    \renewcommand\arraystretch{3} 
    \caption{Accessible datasets for studying the interaction between proteins and RNAs. 
    }
    
   \begin{tabular}	{
   m{1.5cm}<{\centering} || 
   m{2.5cm}<{\centering}
   m{3cm}<{\centering} 
   m{3.5cm}<{\centering} 
   m{5cm}<{\centering}
   }
    \hline
    Type & Dataset Name & Samples &  Availability & Benchmark Methods \\
    \midrule
    \multirow{5}{*}{\makecell[c]{Sequence\\ Dataset}} & \makecell[c]{doRiNA \\ \citep{anders2012dorina}} &
    \makecell[c]{\textcolor{black}{67} RBPs }& \href{https://dorina.mdc-berlin.de/}{https://dorina.mdc-berlin.de/} &
    \makecell[c]{
    iONMF\citep{stravzar2016orthogonal}\\ DeepBind\citep{RN1427}\\ iDeep\citep{RN1412}\\ iDeepS\citep{RN1404}\\ iDeepE\citep{pan2018predicting}\\ GraphProt\citep{RN1433}\\deepnet-rbp\citep{pan2018learning}\\ deepRAM\citep{RN1396}
    }\\
    
    \multirow{5}{*}{}  & \makecell[c]{iCount \\ } &
    \makecell[c]{\textcolor{black}{17} RBPs \\ \textcolor{black}{5,996 binding sites} }&
    \href{https://icount.readthedocs.io/en/latest/index.html}{https://icount.readthedo-cs.io/en/latest/index.html}&
    \makecell[c]{
    iONMF\citep{stravzar2016orthogonal}\\
    iDeepS\citep{RN1404}\\
    iDeepE\citep{pan2018predicting}\\
    deepRAM\citep{RN1396}\\
    } \\
    
    \multirow{5}{*}{}  & \makecell[c]{AURA 2 \\ \citep{dassi2014aura} } &
    \makecell[c]{\textcolor{black}{256} RBPs \\ \textcolor{black}{224,501 binding sites}}&
    \href{http://aura.science.unitn.it/}{http://aura.science.unitn.it/}&
    \makecell[c]{
    RNAcommender\\
    \citep{corrado2016rnacommender}\\
    iDeepE\citep{pan2018predicting}\\
    }\\

   \multirow{5}{*}{}  & 
   \makecell[c]{CLIPdb \\ \citep{yang2015clipdb}}  &
   \makecell[c]{\textcolor{black}{395} CLIP-seq \\ \textcolor{black}{111} RBPs} &
   \href{http://clipdb.ncrnalab.org/}{http://clipdb.ncrnalab.org/}  & 
    deepnet-rbp\citep{pan2018learning}\\

    \hline
    \multirow{2}{*}{\makecell[c]{Protein\\ Structure\\ Dataset}}& \makecell[c]{Protein Data Bank \\ } & 
    \textcolor{black}{179,206} protein structures & 
    \href{https://www.rcsb.org}{https://www.rcsb.org}&
    \makecell[c]{
    NucleicNet \citep{RN1400}\\
    aPRBind \citep{RN1440}\\
    GraphBind \citep{RN1370}\\
    }  \\

    \multirow{2}{*}{} & 
    \makecell[c]{NPIDB} &
    \textcolor{black}{8140} protein structures& 
    \href{https://npidb.belozersky.msu.ru/}{https://npidb.belozers-ky.msu.ru/} &
    NucleicNet \citep{RN1400}\\
    
    \multirow{2}{*}{} & 
    \makecell[c]{AlphaFold DB} &
    \textcolor{black}{23,391 }predicted structures(\textit{Homo sapiens}), all the UniRef90 proteins (over \textcolor{black}{100 million})& 
    \href{https://alphafold.ebi.ac.uk/}{https://alphafold.ebi.ac.uk/} &
    -\\
    
    \hline
    \multirow{2.5}{*}{\makecell[c]{RNA\\ Secondary\\ Structure \\ Dataset}}
    \multirow{5}{*}{}&
    \makecell[c]{bpRNA} & 
    \makecell[c]{\textcolor{black}{102,318} secondary \\ structures} & 
    \href{http://bprna.cgrb.oregonstate.edu/}{http://bprna.cgrb.oregonstate.edu/} &
    - \\

    \multirow{5}{*}{}&
    \makecell[c]{RASP} & 
    \makecell[c]- & 
    \href{http://rasp.zhanglab.net}{http://rasp.zhanglab.net} &
    -
    \\

  \bottomrule
  \end{tabular}
  \label{dataset_table}
\end{table*}

\section{Model inputs and structure encodings}
\label{sec:features}
\begin{figure*}[htp]
  \centering
  \includegraphics[width=0.98\textwidth]{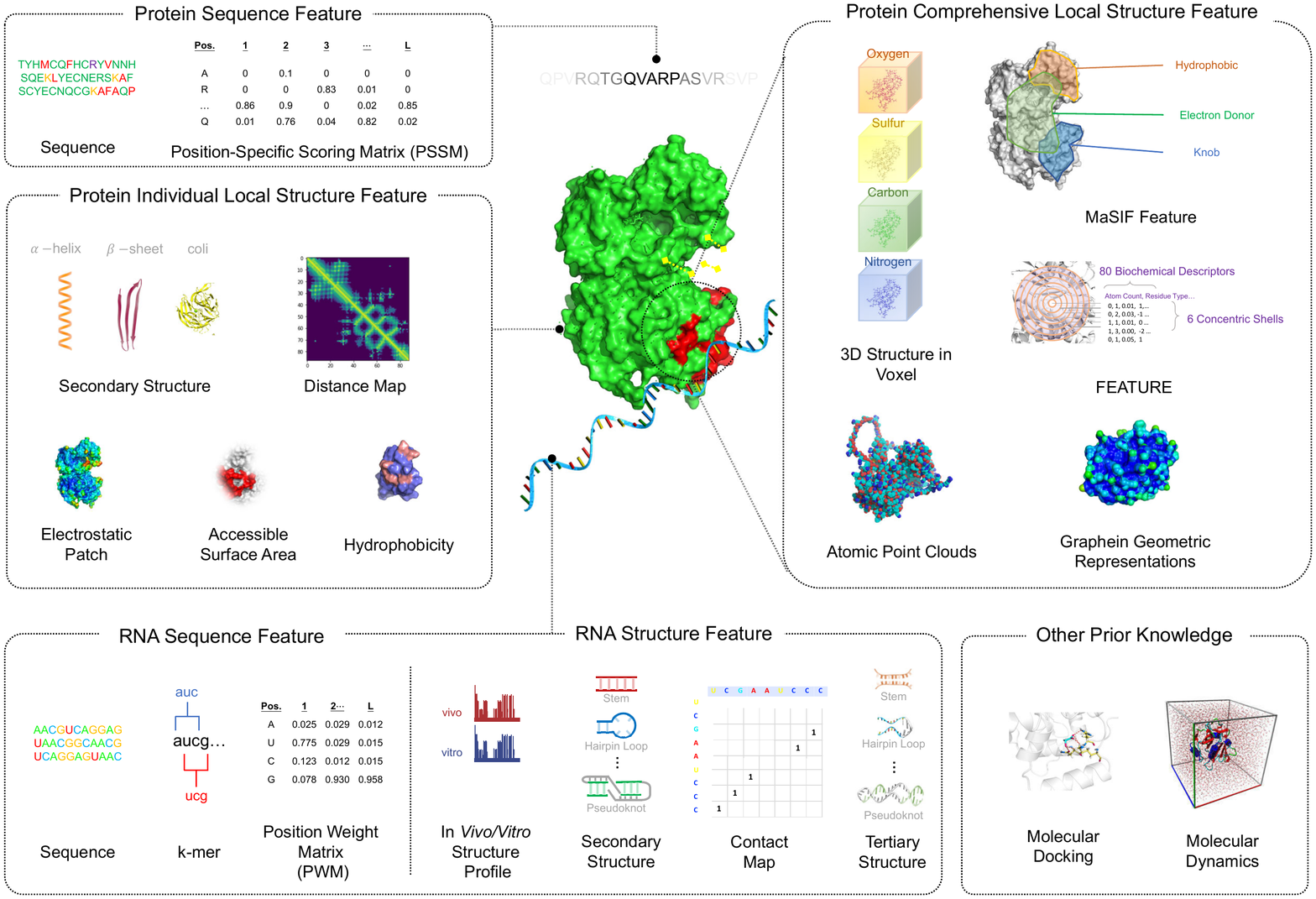}
  \caption{Summary of features from proteins and RNAs, as well as prior knowledge, that can be used to study the interaction between the two molecules.}
\label{figure_feature}
\end{figure*}

The feature and representation of the protein and RNA molecules are crucial for the downstream prediction performance. In this section, we summarize the commonly used encodings of protein and RNA features, including both sequence encoding and structure encoding. We also use Figure \ref{figure_feature} and Table \ref{summary_work} as a summary.

\subsection{\textbf{RNA sequence encodings}}
\textbf{One-hot encoding}: The RNA sequence can be encoded into a 4×L matrix, of which columns correspond to the presence of $A, C, G, U$ and $N$ (padding, if necessary) \citep{xia2019deerect}. Given an RNA sequence $s=(s_{1},s_{2},s_{3}...s_{n})$ with $n$ nucleic acids, and the one-hot encoding matrix M for the sequence is: 
\begin{equation}
M_{i, j}=\left\{\begin{array}{ll}
0.25 & \text { if \ } s_{i}= N, \\
1 & \text { if \ } s_{i}=D_{j}, \\
0 & \text { otherwise, }
\end{array}\right.
\end{equation}
where $i$ is the index of nucleic acids; $D_{j}$ is an ordered list of $[A, C, G, U]$. For the padding sequences, the 4 nucleic acids are assumed to be equally distributed and $[0.25, 0.25, 0.25, 0.25]$ is for the padding nucleotide $N$ in the one-hot matrix.

\textbf{k-mer embedding}: \textcolor{black}{The RNA sequence is split into overlapping $k$-mers \citep{kazan2010rnacontext} of length $k$ using a sliding window with stride $s$. The frequency of each $k$-mer will be directly used as the feature, leading to the loss of contextual information. Subsequently, the word2vec \citep{church2017word2vec} algorithm was applied to extract the additional contextual feature of $k$-mer.}
The word2vec method is an unsupervised learning algorithm that maps $k$-mers from the vocabulary to vectors of real numbers in a low-dimensional space. The embedding representation of $k$-mers is computed in such a way that their context is preserved, \textit{i.e.}, word2vec produces similar embedding vectors for $k$-mers that tend to co-occur or be similar. \textcolor{black}{Generally, the $k$-mer representation is more informative than one-hot encoding \textcolor{black} {\citep{pan2018learning,RN1375}}, while the word2vec algorithm provides contextual information by learning the statistical information of $k$-mer co-occurrence relationships in the input sequences.}

\subsection{\textbf{RNA structure encoding}}
\textbf{RNA secondary structure}: RNA secondary structure offers the local and geometric patterns in two approaches, depending on whether there is an available protein-RNA complexes structure in the PDB. If the structure is available, the explicit secondary structure can be calculated by using an assignment approach, such as RNAstructure \citep{reuter2010rnastructure}. If the structure is unavailable, the predicted secondary structure can be obtained by using a secondary structure prediction algorithm, such as SPOT-RNA \citep{singh2019rna}, RNAshapes \citep{steffen2006rnashapes} and E2Efold \citep{chen2020rna}. For the RNA secondary structure stored in bpRNA \citep{danaee2018bprna}, bpseq file reveals the base pair connection of the RNA.

\textbf{\textit{In vivo} structure profile}: RNA \textit{in vivo} structure profile is produced by \textit{in vivo} click \textcolor{black}{Selective 2'-Hydroxyl Acylation and Profiling Experiment} (icSHAPE) \citep{RN1369}, which is used to characterize the single- and double-stranded regions of RNAs \citep{spitale2015structural}. The raw data of icSHAPE can be processed by the bioinformatic tool, icSHAPE-pipe \citep{li2020icshape}. In brief, raw reads are first collapsed to delete PCR duplicates, and the adapters are trimmed. Next, the clean reads are mapped to the human genome using STAR with the default parameters. Then, icSHAPE scores can be calculated using icSHAPE-pipe, resulting in a $1$×$L$ matrix with the value ranging from 0 to 1.

\textbf{Tertiary structure}: \textcolor{black}{Once given the RNA sequence and the corresponding secondary structural information, JAR3D \citep{roll2016jar3d} can align possible tertiary structural motifs to R3DMA \citep{petrov2013automated}, which contains 253 representative hairpin loop motifs and 276 representative internal loop motifs.} For encoding RNA tertiary structure, the target RNA sequence is first predicted into the probable secondary structure using RNAshapes \citep{steffen2006rnashapes}. Then, all the hairpin and internal loops that overlap the viewpoint region would be fed to JAR3D to calculate the probabilities of folding into the predefined tertiary structural motifs. Subsequently, RNA tertiary structure can be encoded into a binary vector of 529 dimensions, corresponding to 253 hairpin loop motifs and 276 internal loop motifs in the R3MDA. 

\subsection{\textbf{Protein sequence encoding}}
\textbf{One-hot encoding}: The protein sequence can be encoded into a $20$x$L$ matrix, of which columns correspond to the presence of 20 standard amino acids, such as $A, R, N, D.$  The encoding process is similar to that of RNA.

\textcolor{black}{\textbf{Pseudo Amino Acid Composition:} To consider the order information of protein sequence, researchers introduced Pseudo Amino Acid Composition (PseAAC) \citep{chou2001prediction} to represent amino acids composition information and amino acids order information. It is a combination of a set of discrete sequence correlation factors and the 20 components of the conventional amino acid composition.}

\textbf{Position-Specific Scoring Matrix}: \textcolor{black}{PSSM \citep{ahmad2005pssm} introduces evolutionary information into the RNA binding site prediction. PSSM quantifies the conservation of residues, as the binding residues are shown to be conserved in the sequence. The encoding can be conducted by PSI-BLAST \citep{altschul1997gapped}, where the query sequence is aligned to the NCBI non-redundant (nr) sequence database, resulting in a matrix of $20$×$L$. Each value in the matrix represents the frequency of a specific amino acid at a particular position in the multiple sequence alignment \citep{li2018deepre,zou2019mldeepre}.}

\subsection{\textbf{Protein structure encoding}}
\textbf{Local structure}: Individual local structural information included \textcolor{black}{Secondary Structure (SS) \citep{hiller2006using}, Interface Propensity (IP) \citep{li2012new}, Accessible Surface Area (ASA) \citep{heffernan2017capturing}, Electrostatic Patches (EP) \citep{stawiski2003annotating} and \textcolor{black}{Distance Map (DM) \citep{chen2019improve}.}} 
The secondary structure reveals primary structural information, which has 3/8-class labeling systems. Dictionary of Secondary Structure of Protein (DSSP) \citep{kabsch1983dictionary} assigns eight secondary structure states to amino acids, including $3_{10}$-helix G, alpha-helix H, pi-helix I, beta- bridge B, beta-strand E, beta-turn T, and coil C. 
SPIDER3 \citep{heffernan2017capturing} converts the 8-class assignment into the 3-class assignment, where Helix H is composed of G, H, and I; Beta strand B is composed of B and E, Coil C is composed of T and C. 
Li et al. \citep{li2012new} introduces interface propensity, the residue-nucleotide propensities with secondary structure information of proteins and RNAs. 
The propensity of a specific residue-nucleotide pair is calculated from its observed probability at interfaces divided by its expected probability. The interface propensity of a residue type with a particular class of secondary structures is represented as an average value of its pairwise propensities for the four kinds of nucleic acids. 
Accessible surface area is a kind of widely used feature for RNA binding site prediction, which can be calculated by NACCESS \citep{ding2006naccess} when the protein structure is available in PDB. For the protein absent in the PDB, there are several predictive methods, such as ASAquick \citep{faraggi2014accurate} and RNAsol \citep{sun2019enhanced} to predict ASA. 
Electrostatic patches can describe the protein surface charge status, which is an important factor in RNA-binding process. Generally, RNA binding interfaces on protein are more likely to be positively charged, and the electrostatic feature can be calculated by PatchFinderPlus \citep{shazman2007patch}. \textcolor{black}{Distance map can efficiently represent contacted structural information by residue-pairwise distance matrix, which can be calculated by SPOT-Contact \citep{hanson2018accurate}. Distance map has been applied for protein profile prediction, such as solubility \citep{chen2021structure} and drug target interaction \citep{zheng2020predicting}.}

For the comprehensive local structural information, atom features within concentric shells or grid boxes are introduced to describe the physicochemical environment in a specific physical space, which can be calculated by FEATURE \citep{halperin2008feature} or AutoDock \citep{forli2016computational}. 
In FEATURE, 80 physicochemical properties (\textit{e.g.} negative/positive charges, hydrophobicity, solvent accessibility) on atoms of the protein with 7.5Å of a grid point in a radial distribution are divided into six concentric shells of spheres, resulting in a $6$×$80$ matrix. 
AutoDock utilizes an atom-channel (carbon-, oxygen-, nitrogen-, sulfur-) framework to define a local 20Å cubical box to state the presence of carbon, oxygen, sulfur, and nitrogen atoms in a corresponding atom type channel, divided into 1Å cubical voxel, resulting in a $4$×$20$×$20$×$20$ matrix. \textcolor{black}{MaSIF \citep{gainza2020deciphering}, dMaSIF \citep{sverrisson2021fast} and Graphein \citep{jamasb2020graphein} apply geometric deep learning on protein structure surface.}
MaSIF emphasizes the significance of the protein surface, and presents a method to encode geometric features (shape index and distance-dependent curvature) and chemical features (hydropathy, continuum electrostatics and free electrons/protons) on the surface with the geodesic radius of 9Å or 12Å, resulting in a $1$×$80$ matrix. \textcolor{black}{Instead of using surface mesh, dMaSIF employs atomic point cloud representation to extract task-specific geometric and chemical features. Graphein is an efficient tool for constructing graph and surface-mesh representation of protein structures.}

\textbf{Global structure}: Global structural information is rarely used in RNA binding site prediction since the interaction is regarded as a local recognition problem. However, global structural information may play an important role in identifying RBP in future applications. Ishiguro et al. \citep{ishiguro2019graph} introduces supernodes to connect other nodes in the graph representing the compound structure. Proteins with the neighbor-radius contact map could be encoded in a similar way \textcolor{black}{\citep{gligorijevic2021structure}}. 

\section{Computational models}
\label{sec:models}
After encoding the proteins and RNAs, we need to build and train a model to perform the interaction prediction. We divide the methods into two categories, either template-based and shallow-learning methods or deep learning methods, which will be introduced in detail in this section. Table \ref{summary_work} also summarizes the models of different representative works.

\subsection{\textbf{Template-based and shallow-learning models}}
The template-based approach, which is similar to homology modeling, is applied for the binding site prediction with known homologous structures. The models, such as DBD-Threader \citep{gao2009threading}, and SPOT-Seq \citep{singh2019rna}, can directly adopt the known knowledge without feature extraction and mainly rely on the protein structure alignment process. However, for the protein without known homologous structure, the approach is incapable of solving this situation. It is hard for the template-based approach to copy specific sites from few homologous cases. 
On the other hand, the shallow-learning methods attempt to generalize common rules learned from the known experience of a dataset \citep{RN1432}. 
\textcolor{black}{Due to their satisfactory performance and good interpretability, shallow-learning approaches, such as \textcolor{black}{SVM}, random forest, logistic regression, decision tree and naïve Bayes, have been widely used in RNA binding sites and binding preference prediction \citep{zhang2017rbppred,RN1425}.}
Although shallow learning methods are very powerful in terms of interpolation, the prediction of extrapolation can not be guaranteed since the predefined feature limits the module learning from the raw data. The predefined feature provides a explicit but fixed insight of the learning module. However, with the increasing amount of data, the feature extraction procedure can be flexible and learned by the model, so-called deep learning. Generally, deep learning methods yield higher performance of the binding site and binding preference prediction, especially for sophisticated protein \citep{RN1427}. It will be introduced in the following section in detail. On the other hand, several works including RNABindRPlus \citep{walia2014rnabindrplus} and RBRDetector \citep{yang2014rbrdetector}, are attempting to incorporate both template-based and shallow-learning approaches to improve the performance.

\subsection{\textbf{Deep learning models}}
The existing methods emphasize the importance of sequence information. DeepBind \citep{RN1427} is the first deep learning approach for RNA binding preference prediction, which employs a single layer of convolution. DeepBind demonstrates the powerful capability of Convolutional Neural Networks (CNNs) as well as their ability to detect the known motifs. \textcolor{black}{DeepBind} takes only the RNA sequences as inputs and identifies the preference of RNA-binding proteins. Based on DeepBind, DeeperBind \citep{hassanzadeh2016deeperbind} introduces the \textcolor{black}{Long Short-Term Memory (LSTM)} layers into the DeepBind architecture to learn the long-range dependency between the sequence features extracted by the CNN layers. iDeepS \citep{RN1404} also combines CNN and \textcolor{black}{Recurrent Neural Network (RNN)} layers since both of them are helpful for performance, and extra RNA structural motifs are also integrated into the model. iDeepE \citep{pan2018predicting} feeds the local and global sequence information into CNNs, and demonstrates that multiple overlapping fixed-length sub-sequences (similar to $k$-mer) provide informative feature for the binding preference prediction. DeepRAM \citep{RN1396} comprehensively evaluates the model based on CNNs, RNNs, and hybrid CNN/RNN architectures, revealing that the hybrid frameworks outperform the former two architectures. \textcolor{black}{Besides, DeepCLIP \citep{gronning2020deepclip} also employs 1D convolution layers and \textcolor{black}{Bidirectional LSTM (BiLSTM)} to capture the mutation profile of protein-RNA binding preference.} However, single input of sequence limits the model capacity to capture the authentic mechanism of RNA-protein interaction.

With the developing insight of RNA-protein interaction, RNA structural information is discovered to exerts an important role in the binding mechanism. \textcolor{black}{Thus, \textcolor{black}{in order to predict binding site,} deepnet-rbp \citep{RN1426} utilized a multi-modal deep learning framework. It systematically integrated RNA primary sequences, predicted secondary structures from RNAshapes, and tertiary structural features extracted by JAR3D.} \textcolor{black}{As for RNA binding preference prediction, DLPRB \citep{RN1409} also took the advantage of the predicted secondary structures to explore RNA structural contexts.} The PrismNet \citep{RN1369} considered that there are a large number of structurally variable sites across the cell lines. Consequently, icSHAPE \citep{li2020icshape} was introduced in PrismNet to describe the \textit{in vivo} structural profile with $1$×$L$ matrix (see 4.2.2). The PrismNet encodes the sequence with the one-hot encoding and extra \textit{in vivo} structure scores as the fifth dimension. Besides, PrismNet applied a squeeze-and-excitation module \citep{hu2018squeeze} to adaptively calibrate convolutional channels of channel-wise attention and residual blocks, and capture the joint sequence-and-structural binding determinants.

\textcolor{black}{In addition, the protein local structural environment of the binding sites is also crucial to the RNA-protein interaction. Torng and Altman \citep{torng2019high} applied 3D CNNs to protein structure information, generated by AutoDock or FEATURE, and provided the comparable performance as the former RNA-protein interaction binding site prediction method. Furthermore, NucleicNet \citep{RN1400} considered the RNA-binding issue from the perspective of three-dimensional protein structure, which is extracted in units of residues. In order to extract RNA-binding properties in various locations on protein structure, the FEATURE \citep{RN802} framework is used to encode physicochemical properties on the grid point of protein surfaces. For each viewpoint, a high-dimensional feature vector for six concentric shells of spheres with 80 physicochemical properties for each shell will be generated. Furthermore, the NucleicNet predictor used the hierarchical classification of residue sites, first for binding or not, if affirmative, the possible type of RNA constituent binding to the location.} 

To efficiently capture such structural information of RNA and protein local environment, many studies applied Graph Neural Networks (GNNs) to extract the comprehensive features. RPI-Net \citep{RN1375} employed an end-to-end learning approach with GNN from the sequences and structures of RNAs, which provide dense information for binding site prediction. \textcolor{black}{For the graph construction of protein structural context, GraphBind \citep{xia2021graphbind} defined a sliding sphere in the 3D space for the target residue and applied a \textcolor{black}{Hierarchical GNN} (HGNN) to learn the latent patterns of structural and physicochemical characteristics for binding residue recognition.}

\begin{table*}[thp]
    \caption{Summary and comparison of the representative works for studying the protein-RNA interaction. A more comprehensive list is in the \textcolor{black}{Appendix}.
    }
\begin{center}
   \begin{threeparttable}
   \begin{tabular}	{
   m{1cm}<{\centering} 
   m{1cm}<{\centering}
   m{2cm}<{\centering} 
   m{2cm}<{\centering} 
   m{4cm}<{\centering}
   m{5.3cm}<{\centering}
   }
    \hline
    \multirow{2}{*}{Paper} & \multirow{2}{*}{Year} & \multirow{2}{*}{Prediction} & \multirow{2}{*}{Model} & \multicolumn{2}{c}{Feature} \\
    
    \cline{5-6}
     &\multirow{2}{*}{} &\multirow{2}{*}{} &\multirow{2}{*}{} & \textcolor{black}{Feature Encoding Format} & \textcolor{black}{Feature Information} \\
    \hline 
     
    \citep{RN1441} & 2004 & \textcolor{orange}{Binding Site} & Fully-connected NN & \textcolor{black}{Feature vector} & Sequence composition, sequence neighbourhood, SA\tnote{1}  \\
    
    \citep{hiller2006using} & 2006 & \textcolor{blue}{Binding Preference} & PWM & \textcolor{black}{Single-stranded motif finding} & \textcolor{black}{RNA sequence and SS\tnote{2}} \\
    
    \citep{kazan2010rnacontext} & 2010 & \textcolor{blue}{Binding Preference} & PWM & \textcolor{black}{Learning a motif model to build structure annotations} & \textcolor{black}{RNA sequence and SS} \\
    
    \citep{yang2013protein} & 2013 & \textcolor{orange}{Binding Site} & \textcolor{black}{Clustering, maximum voting} & Structure alignment & \textcolor{black}{Binding-specific substructure, sequence profile} \\
    
    \citep{RN1433} & 2014 & \textcolor{blue}{Binding Preference} & SVM & Graph-kernel & \textcolor{black}{RNA sequence and SS} \\
    
    \citep{li2014quantifying} & 2014 & \textcolor{orange}{Binding Site} & ANN & Feature vector & \textcolor{black}{Sequence, evolutionary conservation, surface deformations, SA, side chains} \\    
    
    \citep{RN1427} & 2015 & \textcolor{blue}{Binding Preference} & CNN & One-hot encoding & \textcolor{black}{RNA sequence} \\
    
    \citep{RN1421} & 2016 & \textcolor{blue}{Binding Preference} & \textcolor{black}{PWM} & k-mer embedding & RNA SS \\
    
    \citep{RN1426} & 2016 & \textcolor{blue}{Binding Preference} & Multimodal DBNs & \textcolor{black}{Restricted Boltzmann machines, replicated softmax} & \textcolor{black}{RNA sequence, SS, tertiary structure} \\
     
    \citep{RN789} & 2017 & \textcolor{orange}{Binding Site} & \textcolor{black}{HMM and logistic regression} & \textcolor{black}{PSSM and feature vector} & \textcolor{black}{AA\tnote{3} \ sequence, SS, SA, putative intrinsic disorder and evolutionary information} \\
    
    \citep{jimenez2017deepsite} & 2017 & \textcolor{orange}{Binding Site} & 3D CNN & 3D Voxel & \textcolor{black}{Protein 3D structure with atom properties} \\
    
    \citep{RN1404} & 2018 & \textcolor{blue}{Binding Preference} & CNN+LSTM & One-hot encoding & \textcolor{black}{RNA sequence and SS} \\
    
    \citep{wu2018coach} & 2018 & \textcolor{orange}{Binding Site} & Docking & Structure modeling & \textcolor{black}{Sequence and structure} \\
    
    \citep{pan2018predicting} & 2018 & \textcolor{blue}{Binding Preference}  & Global and local CNN & One-hot encoding & \textcolor{black}{RNA sequence} \\
    
    \citep{RN1399} & 2018 & \textcolor{blue}{Binding Preference} & CNN+RNN & One-hot encoding & \textcolor{black}{RNA sequence and SS} \\

    \citep{torng2019high} & 2019 & \textcolor{orange}{Binding Site} & 3D CNN & \textcolor{black}{3D Voxel} & \textcolor{black}{Atom types, Van der Waals radii} \\
    
    \citep{RN1400} & 2019 & \textcolor{orange}{Binding Site} \textcolor{blue}{\& Preference} & CNN & 
    \textcolor{black}{Feature vector} & \textcolor{black}{Physicochemical characteristics of protein structure surface}  \\
    
    \citep{RN1379} & 2020 & \textcolor{blue}{Binding Preference} & SVM & k-mer embedding & \textcolor{black}{RNA sequence and structure} \\
    
    \citep{RN1370} & 2021 & \textcolor{orange}{Binding Site} & GNN & Graph, feature vector & \textcolor{black}{Pseudo-positions, atomic features, SS, evolutionary conversation} \\

  \bottomrule
  \end{tabular}
    \begin{tablenotes}
    \footnotesize
    \item[1] Solvent accessibility
    \item[2] Secondary structure
    \item[3] Amino acid
  \end{tablenotes}
 \end{threeparttable}
 \label{summary_work}
\end{center}
\end{table*}

\section{Model evaluation}
\label{sec:evaluation}
After building the model, the last step is evaluating the performance of the model to help the users understand the usefulness and weak points of the propose methods. In this section, we summarize the commonly used evaluation criteria in this field.

\subsection{\textbf{Cross-fold and cross-dataset validation}}
Cross-fold (3-, 5-, 10-fold) validation is usually used to evaluate the performance of models with metrics of the \textcolor{black}{Area Under the Receiver Operating Characteristic} (AUROC) and F1 score. For the 10-fold cross-validation, the dataset would be divided into ten folds, and for each time, nine folds of them are used for training while the left one is for testing. 
One problem is that many works are evaluated using data within a specific protein category, indicating that the models only learn protein-specific features instead of general binding features, which limits the application of the models. 
To assess the generalizability of the model, people should also use cross-dataset validation, which means that general models should be established and evaluated with protein data from different categories and different sources \citep{RN1396}.

\subsection{\textbf{Structure visualization}}
The specific patterns inferred from these models can be visualized as the sequence logo diagrams (Weblogo \citep{crooks2004weblogo}) for the RBP. Generally, these patterns can be regarded as the RNA motifs, which can be mapped to the RNA-binding motif dataset, CISBP-RNA (8056 records of RBP binding motifs) \citep{ray2013compendium}. Besides, the RNA binding motifs with particular secondary structures, including stems, multiloops, hairpins, internal loops, and dangling, are prone to access the surface of RBPs. Thus, the structural information extracted from the model can explain their binding tendency. 

\subsection{\textbf{\textit{In vitro} and \textit{in vivo} experimental validation}}
RNAcompete assay (RNAC) \citep{ray2017rnacompete} is a large-scale \textit{in vitro} experiment that uses the epitope-tagged RBP to competitively select RNA sequences from a designed pool. In NucleicNet \citep{RN1400}, the authors obtain 7-mer RNA-binding profiles summarized as a $Z$-score for the individual RNA sequence. The RBPs with both available RNAC data and PDB structure, such as PABPC1, PCBP2, PTBP1, RBFOX1, SNRPA, SRSF2, TARDBP, and U2AF2, are tested. The results suggest that NucleicNet is capable of differentiating between the top and bottom ten sequences indicated by RNAC Z-scores. Thus, RNAC is suitable to evaluate the model performance. \textit{In vivo} experimental validation in PrismNet \citep{RN1369} is to distinguish the relevant affinity of the given RBPs, such as SND1 with specific conformation (hairpin) or single-stranded conformation. With different melting-and-folding treatments to perturb RNA structure without altering the sequence, the authors can obtain two conformations of the given RNA, the one refolding into the hairpin structure and the other retaining single-stranded conformation. PrismNet predicts that a double-stranded binding site for SND1, which is consistent with the \textit{in vivo} affinity experiment. 

\section{Challenges and opportunities}
\label{sec:challenges}
We may encounter several challenges when modeling the interaction between proteins and RNAs. 
In terms of the inputs to the models, we need to think of how to encode structural information more efficiently and even considering the dynamic structural information. 
Regarding the model, we should design novel deep learning models, which can process multi-modality data effectively, including the information from proteins and RNAs, as well as our prior knowledge. 
Furthermore, people also care about the model interpretability, that is, what leads the model to make a specific prediction. 
Revisiting the protein-RNA interaction problem and advancement in the related fields, we may want to resolve some more sophisticated but appealing tasks. For instance, because of the recent breakthrough in the protein structure prediction field, it becomes increasingly possible to perform high-resolution \textit{Ab initio} protein-RNA interaction prediction with only the protein sequence information.
Finally, based on the predicted interaction results, people are also eager to design specific molecules with high binding affinity against the target molecule.
In this section, we discuss the challenges and the potential opportunities in this field in detail. 

\subsection{\textbf{Structure encodings}}
As discussed above, structural information is critical to predicting the protein-RNA interaction accurately. However, how to encode the structural information efficiently remains to be an open question. Because deep learning models are also useful to perform feature selection, when encoding the structural information, we should try to preserve as much raw information as possible, especially the spatial information.

Regarding the protein structure, some traditional ways of encoding, such as 3/8-class protein secondary structure, lose too much raw information. FEATURE \citep{RN1400}, defining shells around a location in the 3D space and summarizing the physicochemical properties within each shell, is another popular method. However, using such an encoding, we cannot differentiate the properties within each shell. In the machine learning field, people usually use 3D voxels, point clouds, and polygon mesh to represent 3D objects. 
3D voxel encoding is similar to the 2D pixel. And it was shown to be better than FEATURE in predicting the functional domain of proteins \citep{torng2019high}. However, because we extend the representation to another axes, we need to design a more efficient algorithm for handling the increasing dimension.
Polygon mesh representation collects vertices, edges, and faces to define the surface of the protein structure. The combination of such a representation and geodesical CNN is shown to extract the fingerprint of the protein surface, which can be used to predict the interaction between different molecules \citep{gainza2020deciphering}.
Point cloud methods sample points from the 3D object, using the coordinates of those points to represent the structure of the object. Although it has not been widely applied in this field, it has shown great power in the computer vision field for 3D object classification and segmentation.

In terms of the RNA structure, people usually use the secondary structure profile to encode them, indicating whether each base is single-strand or double-strand. However, this encoding loses too much information. For example, we would not know which base forms the hydrogen bond with the other specific base. Recently, researchers have shown that predicting the RNA secondary structure by predicting the contact map matrix can boost the performance significantly \citep{chen2020rna}. A similar idea can be applied to the protein-RNA interaction prediction. Meanwhile, using the graph to represent the RNA secondary structure is another natural approach \citep{RN1375}. However, we need to specify which information we want to extract from the graph. \textcolor{black}{In addition, a thermodynamic study revealed the vital role of non-canonical bases in RNA structure formation and stability \citep{jolley2017loss}. For example, in the non-canonical base purine, hydrogen replaces the exocyclic amino group of Adenine. This replacement leads to the Purine-Uracil pair containing only one hydrogen bond instead of two hydrogen bonds in the Adenine-Uracil pair, which could affect the stability of the structure. Ideally, these ubiquitous non-canonical bases should be included in structure encoding.}

Despite the specific encoding we may use from the machine learning field, we still need to consider the chemical background of the problem. The structures in the atom-scale are different from the 3D objects in real life. Although we may use rigid bodies to approximate and model them, they are not rigid bodies. The physicochemical properties \citep{RN1400} should be considered when we design the methods.

\subsection{\textbf{Dynamic structure information}}
Another fundamental property of biomolecules that most machine learning methods fail to consider is their dynamics. As we know, biomolecules are not static, rigid bodies. Every part of the molecule is continuously moving and oscillating in high frequency. The apo protein structures would not stay in the state with the lowest energy all the time. Instead, they may change from one sub-optimal state to another from time to time. When it comes to the interaction between two molecules, such as the interaction between proteins and RNAs, the situation will be even more complex. For example, some molecules, such as Argonaute, need to undergo substantial conformation change to bind to RNA sequences. The other proteins may also have conformation changes once incorporating RNAs. This phenomenon leads to two difficulties when we model the protein-RNA interaction. Firstly, the structure database that we rely on is not perfect for providing the structural information that we need. Simply removing the RNA structure from the protein-RNA complex may not reveal the actual protein apo structure. Secondly, failing to model molecule dynamics may lead to the performance degradation of the machine learning method when we apply the method to real-life problems. To resolve the above challenges, we should use both the PDB structures and the information from \textcolor{black}{Molecular Dynamic} simulation (MD simulation). In practice, we may consider the state of a molecule at each time point as a screenshot. The entire protein dynamics trajectory can be considered as a video. Deep learning techniques to process videos, such as multi-instance learning, would be helpful to resolve this challenge.

\subsection{\textbf{Incorporating prior knowledge}}
In addition to the data, researchers have accumulated expertise and prior knowledge about this problem. For example, we know that Aquifex aeolicus Ribonuclease III (Aa-RNase III) is most likely bind with double-stranded RNAs. Incorporating such knowledge into the machine learning model can further boost the model's prediction performance and usefulness. There are multiple ways to achieve that. We can manipulate the data prepared for training the model by up-sampling the class favored by the prior knowledge. When we train the model, such knowledge could be incorporated into the model implicitly. But we should handle the data carefully to avoid overfitting. On the other hand, we may design a specific machine learning model that explicitly incorporates prior knowledge. For example, by embedding constraint optimization as a module into the deep learning model \citep{chen2020rna}, we can reduce the data size requirement for training a deep learning model.

\subsection{\textbf{Using information from both RNA and protein}}
In the previous studies, when predicting the binding sites on the protein surface, people usually only use the information from the protein. On the other hand, researchers often only use the RNA information when modeling the protein's binding preference to the RNA sequences. Because the interaction is related to both molecules, it is more desirable to consider both when modeling the process. However, as protein and RNA are different molecules, it is not reasonable to use just one deep learning model to process them. Instead, we should use multi-modality models. Essentially, for each molecule, we have a deep learning module to extract features from it. Then, the features can be combined to perform the final prediction. In practice, we may pre-train each module separately first and then fine-tune all the modules together in an end-to-end fashion. By considering the two molecules simultaneously, we do not have to train a model for each protein, and we are more likely to obtain one general model, which deciphers the principle behind protein-RNA interaction.

\subsection{\textbf{Model interpretability for structural modeling}}
It is always difficult to explain deep learning models. For the bio-molecular sequence analysis, after the investigation in the past few years, people have proposed a number of methods to explain the prediction of deep learning models \citep{umarov2019promoter,li2019deep,li2021hmd}. Such explanations converge with the motif discovery techniques before the surge of deep learning. However, for the prediction at the structure level, the explanation is much more difficult. In the structure field, we encounter a serious dilemma between explanation and performance, no matter utilizing deep learning or not. For example, those methods with a strong physicochemical foundation and carefully designed force fields usually have inferior performance compared with the machine learning-based methods. Before the wide usage of deep learning in this field, threading and similarity-based methods are also often used. Although such methods cannot handle queries without homologs, researchers know when they will work and when they will not. However, after deep learning methods are applied to this field, people will use them by default because of their superior performance, although researchers cannot explain what physicochemical and structural biology knowledge are used by the model to perform the prediction. Currently, the request for model interpretability in the structure field is not very urgent because people were still struggling with the performance before the appearance of AlphaFold2 \citep{jumper2021highly}. However, with the fast performance improvement, it is foreseeable that the demand for an explanation of the model will soon increase. The model explanation techniques from the machine learning field can be used to identify which input features influence the final prediction. However, such an explanation is too trivial for this field. Building the connection between the feature and the biological insight would be a more interesting problem, requiring more effort from the researchers.

\subsection{\textbf{High-resolution prediction}}
When predicting the binding sites on the protein surface, researchers usually annotate at the amino acid level. Regarding the binding preference against the RNA sequences, the resolution is usually until the nucleotide. From the structural aspect, the above prediction resolution is still too low. In reality, when studying the interaction between proteins and RNAs, we want to know the exact binding pocket and even the binding location on the protein and RNA surface. With such information, we can understand the functional mechanisms of those important proteins, such as Ago and CRISPR-associated proteins. Some recent works are trying to increase the resolution of the prediction \citep{RN1400,gainza2020deciphering}. More works can be done to improve the existing methods further. For example, although Lam et al. \citep{RN1400} generates grid points on the protein surface and predicts at the grid point level, which increases the prediction resolution significantly on the protein side, the authors have not considered the information from the RNA side at all. Consequently, the method is unable to determine the sequence and orientation of the binding RNA precisely. Introducing features from the RNA structure should increase the prediction resolution for the RNA, although the entire framework needs to be redesigned. 
As discussed in the previous sections, with more advanced structural encoding techniques and frameworks considering both protein and RNA information, the prediction resolution would be increased significantly in the near future.

\subsection{\textbf{\textit{Ab initio} prediction}}
Currently, when predicting the interaction between proteins and RNAs with structural information, people usually assume that we have already known the protein structure. However, in reality, determining the protein and RNA structure is not a trivial task. Even if we can determine the structure of molecules in nature by biological experiments, it is almost impossible to resolve the structure of molecules with mutations, which is important for drug discovery and development. Under that circumstance, it is desirable that we can predict everything from the protein and RNA sequences, which is referred as \textit{Ab initio} prediction here. With the sequences, we may first predict the 3D structures of proteins and RNAs. Then, based on the predicted structures, we will further predict their interactions. Although this research paradigm seems to be computational daunting and may accumulate errors in the multiple steps, it becomes increasingly appealing with the rapid development of the protein structure prediction algorithms in recent years. For example, AlphaFold2 \citep{jumper2021highly} can already achieve a similar prediction accuracy and resolution as Cryo-EM on some proteins. \textcolor{black}{For RNAs, recently proposed deep learning-based method, namely, Atomic Rotationally Equivariant Scorer (ARES) \citep{townshend2021geometric}, has significantly improved prediction of RNA structures.} Eventually, we can use one end-to-end deep learning model to address the two steps all at once. If we could predict the structural interaction details only using the sequence information, gene regulation and drug discovery investigation will be accelerated significantly.

\subsection{\textbf{From prediction to design}}
After determining the molecular structure, we want to know the molecular function, that is, how a specific molecule can interact with another. However, only investigating their function is not our ultimate goal. Eventually, we want to design particular molecules with desirable functions so that to resolve the problems that we encounter in real life, such as curing diseases. As the performance of prediction models has been improved significantly in recent years, researchers are increasingly interested in designing. For instance, people have been using deep learning to optimize the CRISPR guide RNA design \citep{chuai2018deepcrispr,wang2019optimized}. Deep learning has also shown its power in designing new antimicrobial peptide \citep{das2021accelerated}. Regarding this specific topic of protein-RNA interaction, people are especially interested in designing RNA sequences with high binding affinity to protein, similar to the CRISPR guide RNA designing mentioned above. Moreover, a suitable guide RNA for Ago can also increase the gene knock-down efficiency \citep{RN1400}. In addition to the commonly used generative models, such as GAN \citep{goodfellow2014generative} and VAE \citep{kingma2014autoencoding}, recently, differentiable algorithms \citep{chen2020rna} and energy models \citep{song2020score} have drawn great attention in the machine learning field, which is potentially useful for designing problems in the protein-RNA interaction field.

\section{Conclusion}
The interactions between different molecules are essential for biological processes in our body. Among them, the RBP-RNA interactions are of great interest to researchers, considering their central role in gene expression regulation \citep{RN1444,dai2017sequence2vec}. People have developed a number of computational tools and methods to facilitate the study of the RBP-RNA interaction, usually predicting the binding sites and binding preference. However, as we discussed in detail in the review, due to the limitation of the previous data, researchers usually only consider the sequence information and auxiliary structural information to perform the prediction. Considering the recent progress of AlphaFold and the tremendous amount of structure data produced by it \citep{tunyasuvunakool_highly_2021}, the study of the RBP-RNA interactions will be promoted significantly by deep learning methods \citep{RN1400,li2020modern} operating directly on the structure data. 




\begin{appendices}

\onecolumn
\section{Appendix Table}
\begin{ThreePartTable}
\begin{TableNotes}
 \item[1] Solvent accessibility
 \item[2] Secondary structure
 \item[3] Amino acid
\end{TableNotes}
\begin{longtable}{
  m{1cm}<{\centering} 
  m{1cm}<{\centering}
  m{2cm}<{\centering} 
  m{2cm}<{\centering} 
  m{4cm}<{\centering}
  m{5cm}<{\centering}
  }
    \caption{A comprehensive summary and comparison of the representative works for studying the protein-RNA interaction.}
    \label{feature_table_appendix}
    \renewcommand\arraystretch{1.2} \\
    \hline
    \multirow{2}{*}{Paper} & \multirow{2}{*}{Year} & \multirow{2}{*}{Prediction} & \multirow{2}{*}{Model} & \multicolumn{2}{c}{Feature} \\
    \cline{5-6}
    &\multirow{2}{*}{} &\multirow{2}{*}{} &\multirow{2}{*}{} & \textcolor{black}{Feature Encoding Format} & \textcolor{black}{Feature Information} \\
     
    \midrule
    \citep{RN1441} & 2004 & \textcolor{orange}{Binding Site} & Fully-connected NN & \textcolor{black}{Feature vector} & Sequence composition, sequence neighbourhood, SA\tnote{1} \\
    
     \citep{hiller2006using} & 2006 & \textcolor{blue}{Binding Preference} & PWM & \textcolor{black}{Single-stranded motif finding} & \textcolor{black}{RNA sequence and SS\tnote{2}} \\
    
    \citep{kazan2010rnacontext} & 2010 & \textcolor{blue}{Binding Preference} & PWM & \textcolor{black}{Learning a motif model to build structure annotations} & \textcolor{black}{RNA sequence and SS} \\
    
    \citep{yang2013protein} & 2013 & \textcolor{orange}{Binding Site} & \textcolor{black}{Clustering, maximum voting} & Structure alignment & \textcolor{black}{Binding-specific substructure, sequence profile} \\
    
    \citep{li2014quantifying} & 2014 & \textcolor{orange}{Binding Site} & ANN & Feature vector & \textcolor{black}{Sequence, evolutionary conservation, surface deformations, SA, side chains} \\   
    
    \citep{RN1433} & 2014 & \textcolor{blue}{Binding Preference} & SVM & Graph-kernel & \textcolor{black}{RNA sequence and SS} \\
    
    \citep{chen2014identifying} & 2014 & \textcolor{orange}{Binding Site} & \textcolor{black}{Decision tree} & \textcolor{black}{Score ranking} & \textcolor{black}{Electrostatic and evolutionary features of residues} \\
    
    \citep{RN1427} & 2015 & \textcolor{blue}{Binding Preference} & CNN & One-hot encoding & \textcolor{black}{RNA sequence} \\
    
    \citep{RN1421} & 2016 & \textcolor{blue}{Binding Preference} & \textcolor{black}{PWM} & k-mer embedding & RNA SS \\
    
    \citep{zheng2016template} & 2016 & \textcolor{blue}{Binding Preference} & Template & \textcolor{black}{Sequence, structure alignment and transformation matrix} & \textcolor{black}{Sequence and structure} \\
    
    \citep{RN1425} & 2016 & \textcolor{orange}{Binding Site} & Random Forest & Euclidean distance  & \textcolor{black}{Electrostatic feature, triplet interface propensit, PSSM, geometrical and physicochemical properties} \\
    
    \citep{RN1426} & 2016 & \textcolor{blue}{Binding Preference} & Multimodal DBNs & \textcolor{black}{Restricted Boltzmann machines, replicated softmax} & \textcolor{black}{RNA sequence, SS, tertiary Structure} \\
     
    \citep{RN789} & 2017 & \textcolor{orange}{Binding Site} & \textcolor{black}{HMM and logistic regression} & \textcolor{black}{PSSM and feature vector} & \textcolor{black}{AA\tnote{3} \ sequence, SS, SA, putative intrinsic disorder and evolutionary information} \\
    
    \citep{li2017deep} & 2017 & \textcolor{blue}{Binding Preference} & Deep boosting & \textcolor{black}{k-mer embedding} & \textcolor{black}{RNA sequence} \\
    
    \citep{jimenez2017deepsite} & 2017 & \textcolor{orange}{Binding Site} & 3D CNN & 3D Voxel & \textcolor{black}{Protein 3D structure with atom properties} \\
    
    \citep{RN1415} & 2017 & \textcolor{blue}{Binding Preference} & CNN & \textcolor{black}{PSSM and k-mer embedding} & Protein and RNA sequence \\
    
    \citep{zhang2017rbppred} & 2017 & \textcolor{orange}{Binding Site} & SVM & \textcolor{black}{Feature vector} & \textcolor{black}{Physicochemical and evolutionary information of protein sequences} \\
    
    \citep{RN1404} & 2018 & \textcolor{blue}{Binding Preference} & CNN+LSTM & One-hot encoding & \textcolor{black}{RNA sequence and SS} \\
    
    \citep{RN1401} & 2018 & \textcolor{blue}{Binding Preference} & \textcolor{black}{Greedy search} & \textcolor{black}{k-mer embedding} & \textcolor{black}{RNA sequence and structure} \\
    
    \citep{wu2018coach} & 2018 & \textcolor{orange}{Binding Site} & Docking & Structure modeling & \textcolor{black}{Sequence and structure} \\
    
    \citep{pan2018predicting} & 2018 & \textcolor{blue}{Binding Preference}  & Global and local CNN & One-hot encoding & \textcolor{black}{RNA sequence} \\
    
    \citep{wu2018coach} & 2018 & \textcolor{orange}{Binding Site} & CNN & Feature Vector & \textcolor{black}{Hydrophobicity, normalized van der Waals volume, polarity, polarizability, charge and polarity of side chain} \\
    
    \citep{su2019improving} & 2019 & \textcolor{orange}{Binding Site} & SVM & \textcolor{black}{PSSM and feature vector} & \textcolor{black}{Protein sequence and structure} \\
    
    \citep{RN1399} & 2018 & \textcolor{blue}{Binding Preference} & CNN+RNN & One-hot encoding & \textcolor{black}{RNA sequence and SS} \\
    
    \citep{RN1393} & 2019 & \textcolor{blue}{Binding Preference} & CNN & \textcolor{black}{k-mer embedding and one-hot encoding} & \textcolor{black}{RNA sequence and SS} \\

    \citep{torng2019high} & 2019 & \textcolor{orange}{Binding Site} & 3D CNN & \textcolor{black}{3D Voxel} & \textcolor{black}{Atom types, Van der Waals radii} \\
    
    \citep{RN1397} & 2019 & \textcolor{blue}{Binding Preference} & \textcolor{black}{Capsule Network\citep{sabour2017dynamic}} & One-hot encoding & \textcolor{black}{RNA sequence and SS} \\
    
    \citep{RN1400} & 2019 & \textcolor{orange}{Binding Site} \textcolor{blue}{\& Preference} & CNN & \textcolor{black}{Feature vector} & \textcolor{black}{Physicochemical characteristics of protein structure surface}  \\
    
    \citep{RN1372} & 2020 & \textcolor{blue}{Binding Preference} & Recommendation system & \textcolor{black}{FastText\citep{bojanowski2017enriching}} & \textcolor{black}{Protein and RNA sequence} \\
    
    \citep{RN1373} & 2020 & \textcolor{blue}{Binding Preference} & Alignment & \textcolor{black}{PSSM} & \textcolor{black}{RNA sequence and SS} \\
    
    \citep{RN1440} & 2020 & \textcolor{orange}{Binding Site} & CNN & \textcolor{black}{PSSM and feature vector} & \textcolor{black}{Sequence, structure, interface propensity, physicochemical, topology, evolutionary properties and residue fluctuation dynamics} \\
    
    \citep{RN1375} & 2020 & \textcolor{blue}{Binding Preference} & GNN & \textcolor{black}{One-hot encoding, k-mer embedding and PSSM} & \textcolor{black}{RNA sequence and SS} \\
    
    \citep{RN1379} & 2020 & \textcolor{blue}{Binding Preference} & SVM & k-mer embedding & \textcolor{black}{RNA sequence and structure} \\
    
    \citep{gronning2020deepclip} & 2020 & \textcolor{blue}{Binding Preference} & \textcolor{black}{CNN+BiLSTM} & One-hot encoding & \textcolor{black}{RNA Sequence} \\
    
    \citep{RN1370} & 2021 & \textcolor{orange}{Binding Site} & GNN & Graph, feature vector & \textcolor{black}{Pseudo-positions, atomic features, SS, evolutionary conversation} \\
    
    \citep{RN1369} & 2021 & \textcolor{blue}{Binding Preference} & \textcolor{black}{SENet\citep{hu2018squeeze}} & One-hot encoding & RNA sequence and SS \\
    
    \bottomrule
    \insertTableNotes
\end{longtable}
\end{ThreePartTable}

\section{Abbreviation Definitions}
\begin{center}
\begin{table}[thp]
    \caption{\textcolor{black}{A full list of abbreviations in the article and their definitions.} 
    }
    \centering
   \begin{tabular}{ll}
    \hline
Abbreviation & Definition                                                              \\
    \hline 
AA           & Amino Acid                                                              \\
ANN          & Artificial Neural Network                                               \\
AUROC        & Area Under the Receiver Operating Characteristic                        \\
BLSTM        & Bidirectional LSTM                                                      \\
CLIP         & Crosslinking and Immunoprecipitation                                    \\
CNN          & Convolutional Neural Network                                            \\
DBN          & Deep Belief Network                                                     \\
DM           & Distance Map                                                            \\
DSSP         & Dictionary of Secondary Structure of Protein                            \\
DTI          & Drug-Target Interaction                                                 \\
EP           & Electrostatic Patches                                                   \\
GAN          & Generative Adversarial Network                                          \\
GNN          & Graph Neural Network                                                    \\
HGNN         & Hierarchical Graph Neural Network                                       \\
icSHAPE      & \textit{in vivo} click Selective 2'-Hydroxyl Acylation and Profiling Experiment  \\
IP           & Interface Propensity                                                    \\
LSTM         & Long Short-Term Memory                                                  \\
MD           & Molecular Dynamic                                                       \\
MPNPs        & Message Passing Neural Processes                                        \\
NPIDB        & Nucleic Acid–Protein Interaction Database                               \\
PDB          & Protein Data Bank                                                       \\
PseAAC       & Pseudo Amino Acid Composition                                           \\
PSSM         & Position-Specific Scoring Matrix                                        \\
PWM          & Position Weight Matrix                                                  \\
RASP         & RNA Atlas of Structure Probing                                          \\
RBP          & RNA-Binding Protein                                                     \\
RNAC         & RNAcompete assay                                                        \\
RNN          & Recurrent Neural Network                                                \\
RNP          & Ribonucleoprotein                                                       \\
SVM          & Support Vector Machine                                                  \\
UTRs         & Untranslated Regions                                                    \\
VAE          & Variational Auto-Encoder                                                \\
  \bottomrule
  \end{tabular}
 \label{Abbreviation}
\end{table}
\end{center}
\end{appendices}





\bibliographystyle{unsrt}
\bibliography{reference}



\end{document}